\documentclass[proceedings]{JHEP3}

\PrHEP{PrHEP hep2001}                   
\conference{International Europhysics Conference on HEP}                

\usepackage{epsfig}                   

\title{The critical point of lattice QCD on the $\mu$-T plane
}

\author{\speaker{Z. Fodor} 
        \\  
        Institute for Theoretical Physics, Eotvos University,
	Pazmany P. 1, H-1117~Budapest, Hungary\\    
        E-mail: \email{fodor@poe.elte.hu}}                       

\author{S.D. Katz\\                                             
        DESY, Deutsches Elektronen Synchrotron, Theory Group, 
        Notkestrasse 85, D-22603~Hamburg, Germany \\ 
        E-mail: \email{katz@bodri.elte.hu}}

\abstract{We propose a method to study lattice QCD at finite temperature and 
chemical potential.
   We compare it with direct results and with the Glasgow method
      by using $n_f$=4 QCD at Im($\mu$)$\neq$0. We locate the critical
         endpoint (E) of QCD on the Re($\mu$)-T plane. In this study we
	    use $n_f$=2+1 dynamical staggered quarks with semi-realistic
	       masses on $L_t$=4 lattices. }

\begin{document}

QCD at finite $T$ and $\mu$ is of fundamental importance,
since it describes relevant features of particle physics
in the early universe, in neutron stars and in heavy ion collisions.
Extensive experimental work has been done
with heavy ion collisions at CERN and Brookhaven to explore
the $\mu$-$T$ phase diagram.  Note, that
past, present and future heavy ion
experiments with always higher and higher energies produce states
closer and closer to the $T$ axis of the $\mu$-$T$ diagram. It is
a long-standing non-perturbative question, whether a critical point
exists on the $\mu$-$T$ plane,
and particularly how to tell its location theoretically
\cite{crit_point}.           
  
Let us discuss first the $\mu$=0 case.
Universal arguments \cite{PW84} and lattice simulations \cite{U97}
indicate that in a hypothetical QCD
with a strange (s) quark mass ($m_s$) as small as the up (u) and down (d)
quark masses ($m_{u,d}$)
there would be a first order finite
$T$ phase transition. The other extreme case ($n_f$=2)
with light u/d quarks but with an infinitely large $m_s$
there would be no phase transition only an analytical
crossover. Note, that observables change rapidly during a crossover,
but no singularities appear. 
Between the two extremes there is a
critical strange mass ($m_s^c$) at which one has a second order finite
$T$ phase transition. Staggered lattice results on $L_t$=4 lattices
with two light quarks and $m_s$ around the transition $T$ ($n_f$=2+1)
indicated \cite{B90} that $m_s^c$ is about half of the physical $m_s$.
Thus, in the real world we probably have a crossover. 

Returning to a non-vanishing $\mu$, one realizes that arguments
based on a variety of models (see e.g. \cite{B89,qcd_phase,crit_point})
predict a first order finite $T$ phase transition at large $\mu$.
Combining the $\mu=0$ and large $\mu$ informations an interesting
picture emerges on the $\mu$-$T$ plane. For the physical $m_s$
the first order phase transitions at large $\mu$ should be connected
with the crossover on the $\mu=0$ axis. This suggests
that the phase diagram features a critical endpoint $E$ (with
chemical potential $\mu_E$ and temperature $T_E$), at which
the line of first order phase transitions ($\mu>\mu_E$ and $T<T_E$)
ends \cite{crit_point}. At this point the phase transition is of
second order and long wavelength fluctuations appear, which
results in (see e.g. \cite{BPSS01}) consequences, similar to
critical opalescence. Passing close enough to ($\mu_E$,$T_E$)
one expects simultaneous appearance of
signatures 
which exhibit nonmonotonic dependence on the
control parameters \cite{SRS99},
since one can miss the critical point on either of two sides.

The location of E is
an unambiguous, non-perturbative prediction of QCD.
No {\it ab initio}, lattice analysis based on QCD was done to locate
the endpoint.  Crude models
with $m_s=\infty$ were used (e.g. \cite{crit_point})
suggesting that $\mu_E \approx$ 700~MeV, which should be smaller
for finite $m_s$. The goal of our
exploratory work is to show how to locate the endpoint by a lattice
QCD calculation. We use full QCD with dynamical $n_f$=2+1
staggered quarks.
   
QCD at finite $\mu$ can be 
formulated on the lattice \cite{HK83}; however, standard 
Monte-Carlo techniques can not be used. The reason 
is that for Re($\mu$)$\neq$0 the determinant of 
the Euclidean Dirac operator is complex. This fact
spoils any importance sampling method. 

An attractive approach to alleviate the problem 
is the ``Glasgow method'' (see e.g. Ref. \cite{glasgow}) in which the 
partition function ($Z$) is expanded in powers of $\exp(\mu/T)$
by using an ensemble of configurations weighted by the $\mu$=0 action. 
After collecting more than 20 million configurations only unphysical
results were obtained. 
The reason is that the $\mu$=0 ensemble does not overlap enough 
with the finite density states of interest \cite{H01}.

\EPSFIGURE{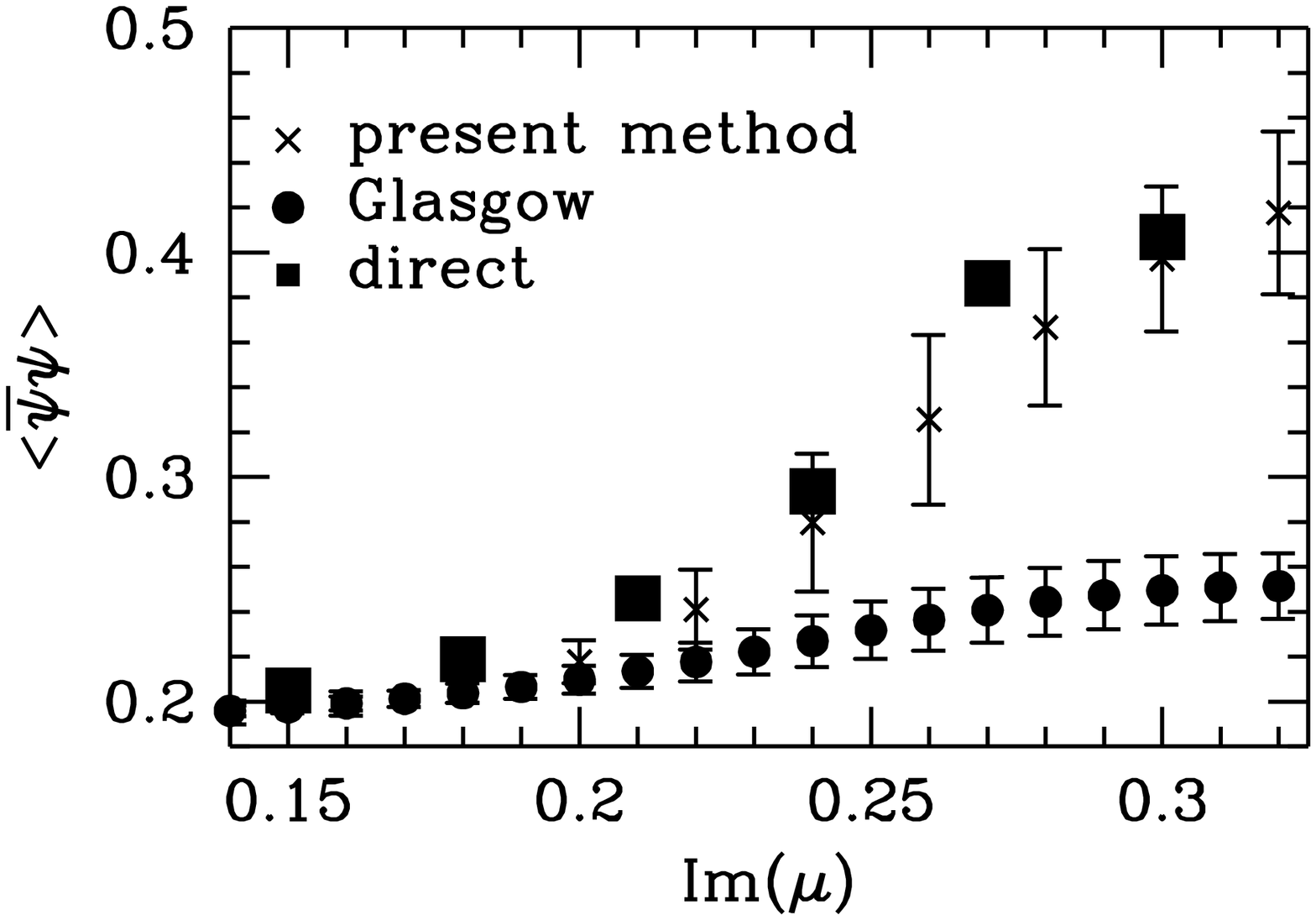,bb= 17 245 570 610, width=7.4cm}
{\it ${\bar \psi}\psi$ as a 
function of Im($\mu$), for direct results (squares),
our technique (crosses) and Glasgow-type reweighting (dots). 
\label{im_mu}}

We propose a method 
to reduce the overlap problem and determine the
phase diagram in the $\mu$-T plane (for details see \cite{FK01}).  
The idea is to produce an ensemble of QCD configurations at
$\mu$=0 and at $T_c$. Then we determine
the Boltzmann weights \cite{FS89} of these configurations at $\mu\neq 0$
and at $T$ lowered to the transition temperatures at this
non-vanishing $\mu$. Since transition configurations
are reweighted to transition configurations a much better
overlap can be observed than by reweighting pure had\-ronic configurations
to transition ones \cite{glasgow}.
After illustrating the applicability of the method 
we locate the critical point of QCD. 
(Multi-dimensional reweighting 
was successful for determining  
the endpoint of the hot electroweak plasma \cite{ewpt}
e.g. on 4D lattices.) 

Let us study a generic system of fermions $\psi$ and bosons $\phi$,
where the fermion Lagrange density is ${\bar \psi}M(\phi)\psi$.
Integrating over the Grassmann fields we get:
\begin{equation}
Z(\alpha)=\int{\cal D}\phi \exp[-S_{bos}(\alpha,\phi)]\det M(\phi,\alpha),
\end{equation}
where $\alpha$ denotes a set of parameters of
the Lagrangian. In the case of staggered QCD $\alpha$
consists of $\beta$,
$m_q$ and $\mu$.
For some choice of the
parameters $\alpha$=$\alpha_0$
importance sampling can be done (e.g. for Re($\mu$)=0).
Rewriting the above equation we get       
\begin{eqnarray}\label{reweight}
Z(\alpha)=
\int {\cal D}\phi \exp[-S_{bos}(\alpha_0,\phi)]\det M(\phi,\alpha_0)&& 
\nonumber \\
\left\{\exp[-S_{bos}(\alpha,\phi)+S_{bos}(\alpha_0,\phi)]
{\det M(\phi,\alpha)  \over \det M(\phi,\alpha_0)}\right\}.&&
\end{eqnarray}
We treat the curly bracket as an observable
(measured on each configuration)
and the rest as the measure. Changing
only one parameter of the ensemble
generated at $\alpha_0$ provides an accurate value for some observables
only for high statistics. This is ensured by 
rare fluctuations as the mismatched measure occasionally sampled the
regions where the integrand is large. This is the 
overlap problem. Having several parameters
the set $\alpha_0$ can be adjusted to get
a better overlap than obtained by varying only one parameter. 

The basic idea of the method as applied to dynamical QCD can be 
summarized as follows. We study the system at ${\rm Re}(\mu)$=0 around 
its transition point. Using a Glasgow-type technique we calculate the 
determinants for each configuration for a set of $\mu$, which, similarly 
to the Ferrenberg-Swendsen method \cite{FS89}, can be used for 
reweighting.  The average plaquette values can be used to perform an 
additional reweighting in $\beta$.  Since transition configurations were 
reweighted to transition ones a much better overlap can be 
observed than by reweighting pure hadronic configurations to transition 
ones as done by the Glasgow-type techniques (moving along the transition 
line was also suggested by Ref. \cite{AKW99}).

We have directly tested these ideas in $n_f$=4 QCD 
with $m_q$=0.05 dynamical 
staggered quarks. 
We first collected 1200 independent V=4$\cdot 6^3$ configurations at 
Re($\mu$)=Im($\mu$)=0 and some $\beta$ 
values and used the Glasgow-reweighting and 
also our technique to study Re($\mu$)=0, Im($\mu$)$\neq$0. At 
Re($\mu$)=0, Im($\mu$)$\neq$0 direct simulations are possible. 
After performing these direct simulations as well, a clear 
comparison can be done. Figure 1 shows the predictions of 
the three methods for the average quark condensates at $\beta$=5.085
as a function of Im($\mu$).
The predictions of our method agree with the direct results,
whereas for larger Im($\mu$) the predictions of the Glasgow
method are by several standard deviations off. 
Based on these experiences we expect that our
method can be applied at Re($\mu$)$\neq$0. 

In QCD with $n_f$ staggered quarks
one should change the determinants to their $n_f$/4 power in our two 
equations. Importance sampling works also in this case  at some $\beta$ and 
at Re($\mu$)=0. Since $\det M$ is complex
an additional problem arises, one should
choose among the possible Riemann-sheets of the fractional power
in eq. (\ref{reweight}). This can be done by using
the fact that at $\mu=\mu_w$ the ratio of the determinants is 1 and
the ratio should be a continuous function of $\mu$.
However, the continuity can only be ensured if the analytical
dependence of the determinant on $\mu$ is known 
\cite{FK01} (the idea goes back to a method of \cite{T90}). 

\EPSFIGURE{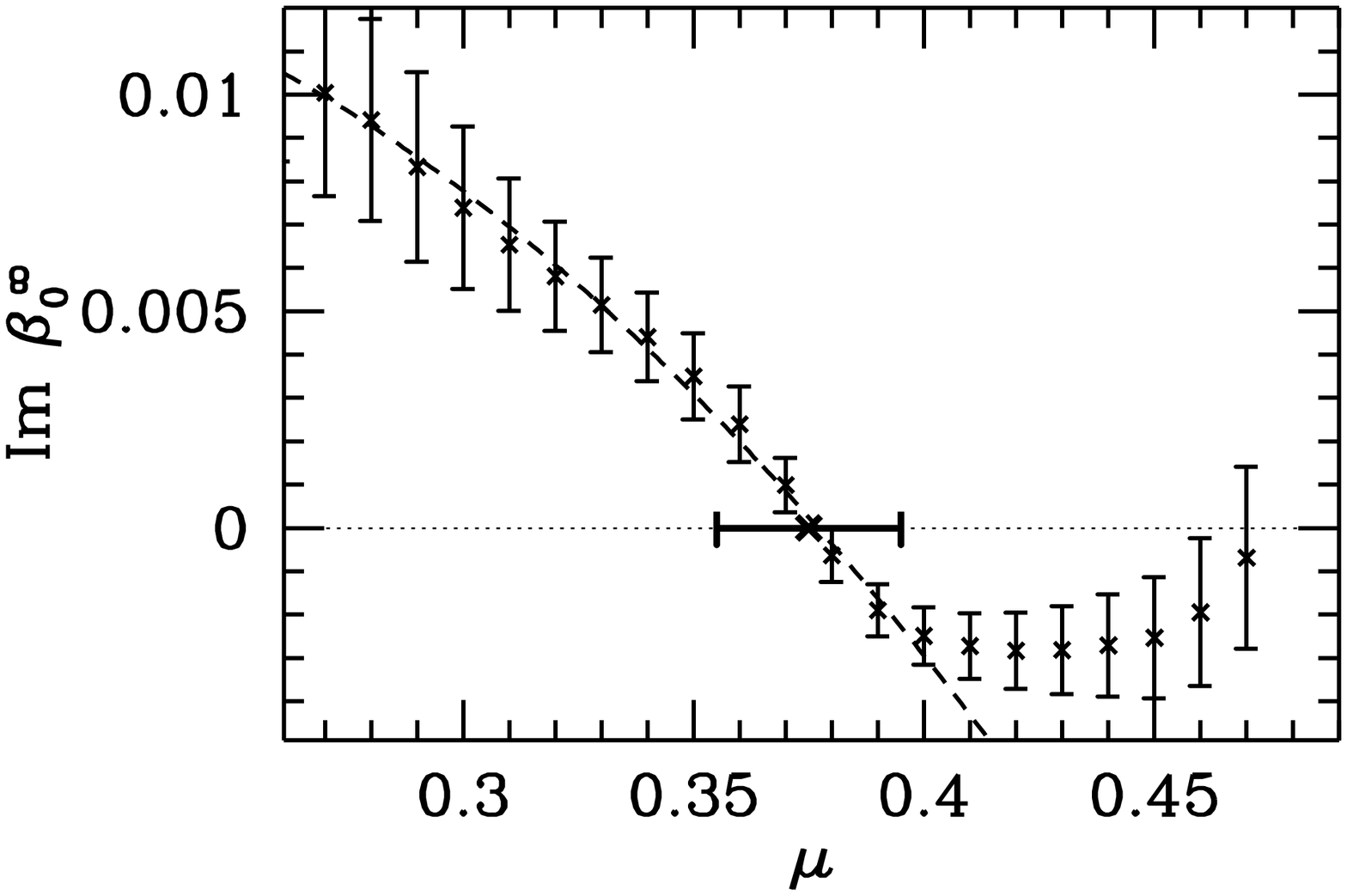,bb= 17 257 570 610,width=7.4cm}
{\it Im($\beta_0^\infty$) as a function of the chemical potential.\label{infV}}

In the 
following we keep $\mu$ real and look for the zeros of $Z$  
for complex $\beta$.  At a first order phase transition the free 
energy $\propto \log Z(\beta)$ is non-analytic. 
A phase transition appears only in the V$\rightarrow \infty$ limit, 
but not in a finite $V$. Nevertheless, $Z$
has zeros at finite V, generating the non-analyticity of the 
free energy, the Lee-Yang zeros \cite{LY52}. 
These are at complex values
of the parameters, in our case at complex $\beta$. For a 
system with a first order transition these zeros
approach the real axis in the V$\rightarrow \infty$ limit
(detailed analysis suggests $1/V$ scaling).   
This V$\rightarrow \infty$ limit generates the non-analyticity of
the free energy. For a system with crossover  
$Z$ is analytic, and the zeros do
not approach the real axis in the V$\rightarrow \infty$ limit.

{\EPSFIGURE{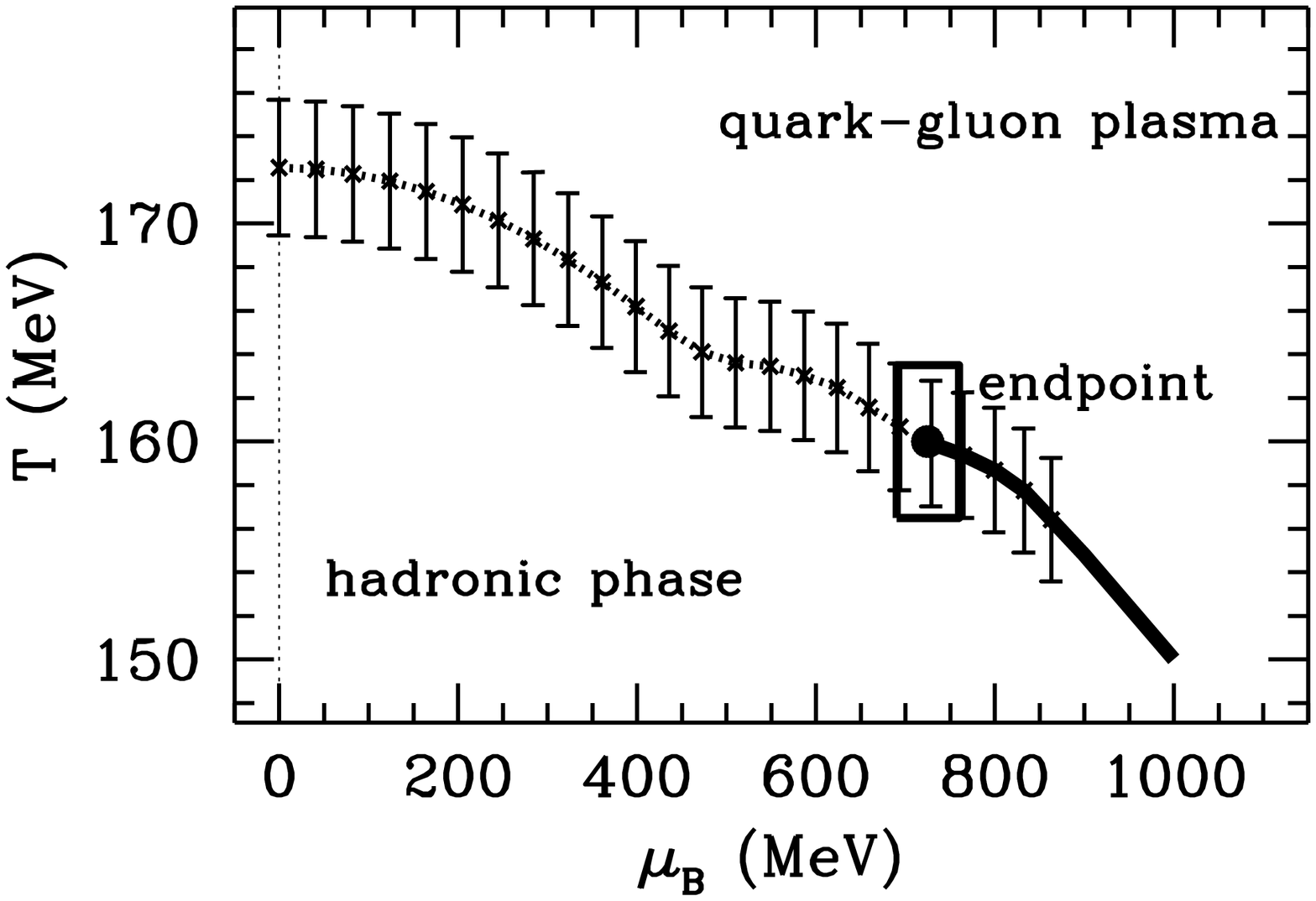,bb= 17 257 570 610,width=7.4cm}
{\it The T-$\mu$ diagram. Direct results are given with errorbars.
Dotted line shows the crossover, solid line the first order 
transition. The box gives the uncertainties of the endpoint.\label{physical}}

At T$\neq$0 we used $L_t$=4, $L_s$=4,6,8 lattices. T=0 runs were done on
$10^3\cdot$ 16 lattices. $m_{u,d}$=0.025 and $m_s$=0.2 were
our bare quark masses.  
At  $T\neq 0$ we determined the complex valued Lee-Yang zeros, 
$\beta_0$, for different V-s as a function of $\mu$. Their 
V$\rightarrow \infty$ limit was given by a $\beta_0(V)=\beta_0^\infty+\zeta/V$
extrapolation. We used 14000, 3600 and 840 configurations on 
$L_s$=4,6 and $8$ lattices, respectively.  
Im($\beta_0^\infty$) is shown on Figure 2 as a function of $\mu$.  For small
$\mu$-s the extrapolated Im($\beta_0^\infty$) is inconsistent with
a vanishing value, and predicts a crossover.
Increasing $\mu$ the value of Im($\beta_0^\infty$) decreases, 
thus the transition becomes consistent with a first order phase
transition. (Note, that systematic overshooting is a finite V effect.)
The statistical error was determined by jackknife samples of the
total $L_s=4,6$ and $8$ ensembles. 
Our primary result is $\mu_{end}=0.375(20)$. 

To set the physical scale we used a 
weighted average of $R_0$,  $m_\rho$  and 
$\sqrt{\sigma}$. 
Note, that (including systematics due to 
finite V) we have 
$(R_0\cdot m_\pi)=0.73(6)$, which is at least twice, $m_{u,d}$ is
at least four times
as large as the physical values. 

Figure 3 shows the phase diagram in
physical units, thus
$T$ as a function of $\mu_B$, the baryonic chemical potential 
(which is three times larger then the quark chemical potential). 
The endpoint
is at $T_E=160 \pm 3.5$~MeV, $\mu_E=725 \pm 35$~MeV.
At $\mu_B$=0 we obtained $T_c=172 \pm 3$~MeV. 

We proposed a method --an overlap improving multi-parameter reweighting 
technique-- to numerically study non-zero $\mu$ and determine the 
phase diagram in the $T$-$\mu$ plane.   
Our method is applicable to any number of Wilson or staggered quarks. 
As a direct test we showed that for Im($\mu$)$\neq$0 the predictions 
of our method are 
in complete agreement with the direct simulations, whereas the Glasgow
method suffers from the well-known overlap problem.
We studied the $\mu$-$T$ phase diagram of QCD with 
dynamical $n_f$=2+1 quarks. 
Using our method we obtained 
$T_E$=160$\pm$3.5~MeV and $\mu_E$=725$\pm$35~MeV for the endpoint. 
Though $\mu_E$ is too
large to be studied at RHIC or LHC, the endpoint would 
probably move closer to the $\mu$=0 axis 
when the quark masses get reduced. 
At $\mu$=0 we obtained $T_c$=172$\pm$3~MeV.
Clearly, more work is needed to get
the final values. One has to  extrapolate 
in the R-algorithm and to the thermodynamic, chiral and continuum limits.

This work was partially supported by Hung. Sci. 
grants No. 
OTKA-\-T34980/\-T29803/\-T22929/\-M28413/\-OM-MU-708/\-IKTA111/\-NIIF. 
This work was in part based 
on the MILC collaboration's public lattice gauge theory code:
http://physics.indiana.edu/\~{ }sg/milc.html.


\begin{thebibliography}{99}
\bibitem{crit_point} M. Halasz {\it et al.} Phys. Rev. D58 (1998) 096007; 
J. Berges, K. Rajagopal, Nucl. Phys. B538 (1999) 215. 
M. Stephanov, K. Rajagopal, E. Shuryak, Phys. Rev. Lett. 81 (1998) 4816; 
T.M. Schwarz, S.P. Klevansky, G. Papp,
Phys. Rev. C60 (1999) 055205.
\bibitem{PW84} R. Pisarski and F. Wilczek, Phys. Rev. D29 (1984) 338;
F. Wilczek, Int. J. Mod. Phys. A7 (1992) 3911; K. Rajagopal and
F. Wilczek, Nucl. Phys. B399 (1993) 395.
\bibitem{U97} For recent reviews see: A. Ukawa, Nucl. Phys. Proc. Suppl.
53 (1997) 106; E. Laerman, {\it ibid}. 63 (1998) 114;
F. Karsch, {\it ibid}. 83 (2000) 14; S. Ejiri,
{\it ibid}. 94 (2001) 19.
\bibitem{B90} F.K. Brown, Phys. Rev. Lett. 65 (1990) 2491;
S. Aoki et al., Nucl. Phys. Proc. Suppl. 73 (1999) 459.
\bibitem{B89} A. Barducci et al., Phys. Lett. B231 (1989) 463;
Phys. Rev. D41 (1990) 1610; {\it ibid}. D49 (1994) 426; S.P. Klevansky,
Rev. Mod. Phys. 64 (1992) 649; M. Stephanov, Phys. Rev. Lett. 76,
(1996) 4472.
\bibitem{qcd_phase} M. Alford, K. Rajagopal and F. Wilczek, Phys. Lett.
B422 (1998) 247; Nucl. Phys. B537 (1999) 443; R. Rapp, T. Sch\"afer,
E.V. Shuryak and M. Velkovsky, Phys. Rev. Lett. 81 (1998) 53; for a
recent review with references see K. Rajagopal and F. Wilczek,
hep-ph/0011333.
\bibitem{BPSS01} S. Bors\'anyi et al., Phys. Rev. D64 (2001) 125011.
\bibitem{SRS99} M. Stephanov, K. Rajagopal and E. Shuryak,
Phys. Rev. D60 (1999) 114028.
\bibitem{HK83} P. Hasenfratz and F. Karsch, Phys. Lett. B125 (1983) 308;
J. Kogut et al., Nucl. Phys. B225 (1983) 93.
\bibitem{glasgow} I.M. Barbour et al., Nucl. Phys. B (Proc. Suppl.)
60A (1998) 220.
\bibitem{H01} For a recent review on QCD at finite T and $\mu$ see e.g. 
S. Hands, hep-lat/0109034.
\bibitem{FK01} Z. Fodor and S.D. Katz, hep-lat/0104001; hep-lat/0106002.
\bibitem{ewpt} Y. Aoki et al., Phys. Rev. D60 (1999) 013001; 
F. Csikor, Z. Fodor and J. Heitger, Phys. Rev. Lett. 82 (1999) 21.
\bibitem{FS89} A.M. Ferrenberg, R.H. Swendsen, Phys. Rev. Lett.
63 (1989) 1195; 61 (1988) 2635.
\bibitem{LY52} C.N. Yang and T.D. Lee, Phys. Rev. 87 (1952) 404.
\bibitem{T90} D. Toussaint, Nucl. Phys. B (Proc. Suppl.) 17 (1990) 
248.
\bibitem{AKW99} M. Alford, A. Kapustin and F. Wilczek, Phys. Rev.
D59 (1999) 054502.
\end{thebibliography}
\end{document}